\documentclass[aps,prl,superscriptaddress,preprint,tightenlines,floatfix,showpacs]{revtex4}

\usepackage[dvips]{graphicx}
\usepackage[dvips]{color}
\usepackage{epsfig}
\usepackage{amssymb,amsmath}
\usepackage{lscape}

\newcommand{\bdk}{$B^{\pm}\to DK^{\pm}$}

\newcommand{\bdsk}{$B^{\pm}\to D^{*}K^{\pm}$}

\newcommand{\bdks}{$B^{\pm}\to DK^{*\pm}$}
\newcommand{\bdtks}{$B^{\pm}\to \tilde{D}K^{*\pm}$}
\newcommand{\bdksnr}{$B^{\pm}\to DK_S\pi^{\pm}$}

\newcommand{\bddsk}{$B^{\pm}\to D^{(*)}K^{\pm}$}

\newcommand{\bdpi}{$B^{\pm}\to D\pi^{\pm}$}

\newcommand{\bddspi}{$B^{\pm}\to D^{(*)}\pi^{\pm}$}

\newcommand{\dsdpi}{$D^{*\pm}\to D\pi^{\pm}$}

\newcommand{\dkpp}{$\bar{D^0}\to K_S\pi^+\pi^-$}
\newcommand{\dtkpp}{$\tilde{D}\to K_S\pi^+\pi^-$}

\begin{document}

\preprint{BELLE-CONF-0502}

\title{Measurement of \boldmath{$\phi_3$} with Dalitz Plot
Analysis of \boldmath{\bdks} Decay}
\affiliation{Aomori University, Aomori}
\affiliation{Budker Institute of Nuclear Physics, Novosibirsk}
\affiliation{Chiba University, Chiba}
\affiliation{Chonnam National University, Kwangju}
\affiliation{Chuo University, Tokyo}
\affiliation{University of Cincinnati, Cincinnati, Ohio 45221}
\affiliation{University of Frankfurt, Frankfurt}
\affiliation{Gyeongsang National University, Chinju}
\affiliation{University of Hawaii, Honolulu, Hawaii 96822}
\affiliation{High Energy Accelerator Research Organization (KEK), Tsukuba}
\affiliation{Hiroshima Institute of Technology, Hiroshima}
\affiliation{Institute of High Energy Physics, Chinese Academy of Sciences, Beijing}
\affiliation{Institute of High Energy Physics, Vienna}
\affiliation{Institute for Theoretical and Experimental Physics, Moscow}
\affiliation{J. Stefan Institute, Ljubljana}
\affiliation{Kanagawa University, Yokohama}
\affiliation{Korea University, Seoul}
\affiliation{Kyoto University, Kyoto}
\affiliation{Kyungpook National University, Taegu}
\affiliation{Swiss Federal Institute of Technology of Lausanne, EPFL, Lausanne}
\affiliation{University of Ljubljana, Ljubljana}
\affiliation{University of Maribor, Maribor}
\affiliation{University of Melbourne, Victoria}
\affiliation{Nagoya University, Nagoya}
\affiliation{Nara Women's University, Nara}
\affiliation{National Central University, Chung-li}
\affiliation{National Kaohsiung Normal University, Kaohsiung}
\affiliation{National United University, Miao Li}
\affiliation{Department of Physics, National Taiwan University, Taipei}
\affiliation{H. Niewodniczanski Institute of Nuclear Physics, Krakow}
\affiliation{Nihon Dental College, Niigata}
\affiliation{Niigata University, Niigata}
\affiliation{Osaka City University, Osaka}
\affiliation{Osaka University, Osaka}
\affiliation{Panjab University, Chandigarh}
\affiliation{Peking University, Beijing}
\affiliation{Princeton University, Princeton, New Jersey 08545}
\affiliation{RIKEN BNL Research Center, Upton, New York 11973}
\affiliation{Saga University, Saga}
\affiliation{University of Science and Technology of China, Hefei}
\affiliation{Seoul National University, Seoul}
\affiliation{Sungkyunkwan University, Suwon}
\affiliation{University of Sydney, Sydney NSW}
\affiliation{Tata Institute of Fundamental Research, Bombay}
\affiliation{Toho University, Funabashi}
\affiliation{Tohoku Gakuin University, Tagajo}
\affiliation{Tohoku University, Sendai}
\affiliation{Department of Physics, University of Tokyo, Tokyo}
\affiliation{Tokyo Institute of Technology, Tokyo}
\affiliation{Tokyo Metropolitan University, Tokyo}
\affiliation{Tokyo University of Agriculture and Technology, Tokyo}
\affiliation{Toyama National College of Maritime Technology, Toyama}
\affiliation{University of Tsukuba, Tsukuba}
\affiliation{Utkal University, Bhubaneswer}
\affiliation{Virginia Polytechnic Institute and State University, Blacksburg, Virginia 24061}
\affiliation{Yonsei University, Seoul}
  \author{K.~Abe}\affiliation{High Energy Accelerator Research Organization (KEK), Tsukuba} 
  \author{K.~Abe}\affiliation{Tohoku Gakuin University, Tagajo} 
  \author{N.~Abe}\affiliation{Tokyo Institute of Technology, Tokyo} 
  \author{I.~Adachi}\affiliation{High Energy Accelerator Research Organization (KEK), Tsukuba} 
  \author{H.~Aihara}\affiliation{Department of Physics, University of Tokyo, Tokyo} 
  \author{M.~Akatsu}\affiliation{Nagoya University, Nagoya} 
  \author{Y.~Asano}\affiliation{University of Tsukuba, Tsukuba} 
  \author{T.~Aso}\affiliation{Toyama National College of Maritime Technology, Toyama} 
  \author{V.~Aulchenko}\affiliation{Budker Institute of Nuclear Physics, Novosibirsk} 
  \author{T.~Aushev}\affiliation{Institute for Theoretical and Experimental Physics, Moscow} 
% ...
\collaboration{The Belle Collaboration}

\date{\today}

\begin{abstract} 
We present a measurement of the unitarity triangle angle $\phi_3$
using a Dalitz plot analysis of the three-body decay of the neutral 
$D$ meson from the \bdks\ process. 
Using a 253 fb$^{-1}$ data sample collected by the Belle experiment,  
we obtain 56 signal candidates for \bdks\ where the neutral 
$D$ meson decays into $K_S \pi^+ \pi^-$.
From a maximum likelihood fit we obtain 
$\phi_3=112^{\circ}\pm 35^{\circ}
\mbox{(stat)}\pm 9^{\circ}
\mbox{(syst)}\pm 11^{\circ}(\mbox{model})\pm 8^{\circ}(\mbox{nonresonant }
B^{\pm}\to DK_S\pi^{\pm})$. 
\end{abstract}
\maketitle

\section{Introduction}

Determinations of the Cabbibo-Kobayashi-Maskawa
(CKM) \cite{ckm} matrix elements provide important checks on
the consistency of the Standard Model and ways to search
for new physics.  Various methods using $CP$ violation in $B\to D K$ decays have been
proposed \cite{glw,dunietz,eilam,ads} to measure the unitarity triangle
angle $\phi_3$. The sensitivity to the angle $\phi_3$ comes 
from the interference of two amplitudes producing opposite 
flavors of neutral $D$ meson. Such a mixed state of neutral $D$ 
will be called $\tilde{D}$. 
For example, in the case of $B^+$ decay, the mixed state is 
$\tilde{D}_+=\bar{D}^0+re^{i\theta_+}D^0$. Here $r$ is the ratio of the 
suppressed and favored amplitudes (it is expected to be of order of 0.1--0.3), 
the total phase between the $\bar{D}^0$ and $D^0$ is $\theta_+=\phi_3+\delta$, 
where $\delta$ is the strong 
phase difference between suppressed and favored decays. 
Analogously, for the decay of $B^-$, one can write
$\tilde{D}_-=D^0+re^{i\theta_-}\bar{D}^0$ with $\theta_-=-\phi_3+\delta$. 

Three body final states common to $D^0$ and
$\bar{D^0}$, such as $K_S\pi^+\pi^-$ \cite{giri}, have been suggested as 
promising modes for the extraction of $\phi_3$. 
The Dalitz plot density of $\tilde{D}$ gives immediate information 
about $r$ and $\theta_\pm$, once the amplitude of the $\bar{D^0}$ decay is known. 
The amplitude of the $\tilde{D}_+$ decay 
as a function of Dalitz plot variables $m^2_+=m^2_{K_S\pi^+}$ and 
$m^2_-=m^2_{K_S\pi^-}$ is 
\[
  M_+=f(m^2_+, m^2_-)+re^{i\phi_3+i\delta}f(m^2_-, m^2_+), 
\]
where $f(m^2_+, m^2_-)$ is an amplitude of the \dkpp\ decay. 
Similarly, 
\[
  M_-=f(m^2_-, m^2_+)+re^{-i\phi_3+i\delta}f(m^2_+, m^2_-). 
\]
The \dkpp\ decay model can be determined
from a large sample of flavor-tagged \dkpp\ decays 
produced in continuum $e^+e^-$ annihilation. Once that is known, 
a simultaneous fit of $B^+$ and $B^-$ data allows to separate the 
contributions of $r$, $\phi_3$ and $\delta$. 
Refer to \cite{giri,belle_phi3} for a more detailed description of 
the technique. 

The method described can be applied to other modes as well as \bdk\ decay.
Excited states of neutral $D$ and $K$ mesons can also be used, although 
the values of $\delta$ and $r$ can differ for these decays. 
Previously Belle~\cite{belle_phi3,belle_new} and Babar~\cite{babar_phi3} 
collaborations
have performed the analyses using this technique involving \bdk\ and 
\bdsk\ decays. This paper describes the analysis using \bdks\ decay. 

\section{Event selection}

We use a 253 fb$^{-1}$ data sample, corresponding to
$275\times 10^6$ $B\bar{B}$ pairs, collected by 
the Belle detector \cite{belle}. The decay chain \bdks\ with 
$D\to K_S\pi^+\pi^-$ and $K^{*\pm}\to K_S\pi^{\pm}$ is 
selected for the analysis. 

The requirements on the quality of the charged tracks are the same
as in our previous analysis \cite{belle_new}. To select $K^{*\pm}$ mesons, 
we require the invariant mass of $K_S\pi^{\pm}$ to be within 
50 MeV/$c^2$ of its nominal mass. For the selection of neutral $D$
meson, the invariant mass $M_{K_S\pi\pi}$ of its decay products
is required to satisfy $|M_{K_S\pi\pi}-m_D|<15$ MeV/$c^2$. 

The selection of $B$ candidates is based on the center-of-mass (CM) energy 
difference
$\Delta E = \sum E_i - E_{\rm beam}$ and the beam-constrained $B$ meson mass
$M_{\rm bc} = \sqrt{E_{\rm beam}^2 - (\sum p_i)^2}$, where $E_{\rm beam}$ 
is the CM beam 
energy, and $E_i$ and $p_i$ are the CM energies and momenta of the
$B$ candidate decay products. The requirements for signal 
candidates are $5.272$~GeV/$c^2<M_{\rm bc}<5.288$ GeV/$c^2$ and $|\Delta E|<0.022$ GeV. 

To suppress background from $e^+e^-\to q\bar{q}$ ($q=u, d, s, c$) 
continuum events, we require $|\cos\theta_{\rm thr}|<0.8$, 
where $\theta_{\rm thr}$ is the angle between the thrust axis of 
the $B$ candidate daughters and that of the rest of the event. 
We also require the helicity angle of $K^{*\pm}$ to be 
$|\cos\theta_{\rm hel}|>0.4$. For additional background rejection, we 
use a Fisher discriminant composed of 11 parameters \cite{fisher}: 
the production angle of the $B$ candidate, the angle of the $B$ thrust 
axis relative to the beam axis and nine parameters representing 
the momentum flow in the event relative to the $B$ thrust axis in the CM frame.
We apply a requirement on the Fisher 
discriminant that retains 95\% of the signal and rejects 30\% of the 
remaining continuum background. 

\begin{figure}[!htb]
  \epsfig{figure=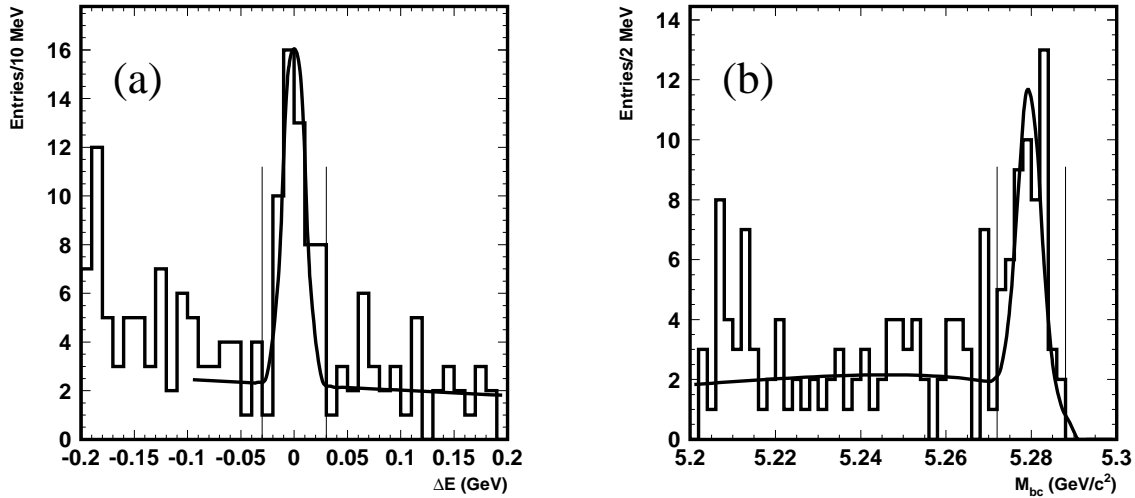,width=\textwidth}
  \vspace{-1\baselineskip}
  \caption{(a) $\Delta E$ and (b) $M_{\rm bc}$ distributions for the \bdks\
  candidates. Vertical lines show the signal region. 
  The histogram shows the data; the smooth curves are the fit result.}
  \label{b2dks_sel}
\end{figure}

The \bdks\ selection efficiency (4.1\%) is determined from 
a Monte Carlo (MC) simulation. The number of events passing all selection 
criteria is 56. 
The $\Delta E$ and $M_{\rm bc}$ distributions for \bdks\ candidates are
shown in Fig.~\ref{b2dks_sel}.
The background fraction is determined from a binned fit to the $\Delta E$
distribution, in which the signal is represented by a Gaussian distribution 
with mean
$\Delta E=0$ and the background is modeled by a 
linear function. The contributions in the signal region are found 
to be $36\pm 7$ signal events and $13\pm 2$ background events. 
The overall background fraction is $27\pm 5$\%.

\section{Dalitz plot analysis of $B^{\pm} \to D K^{*\pm}$ decay}

\begin{figure}[!htb]
  \epsfig{figure=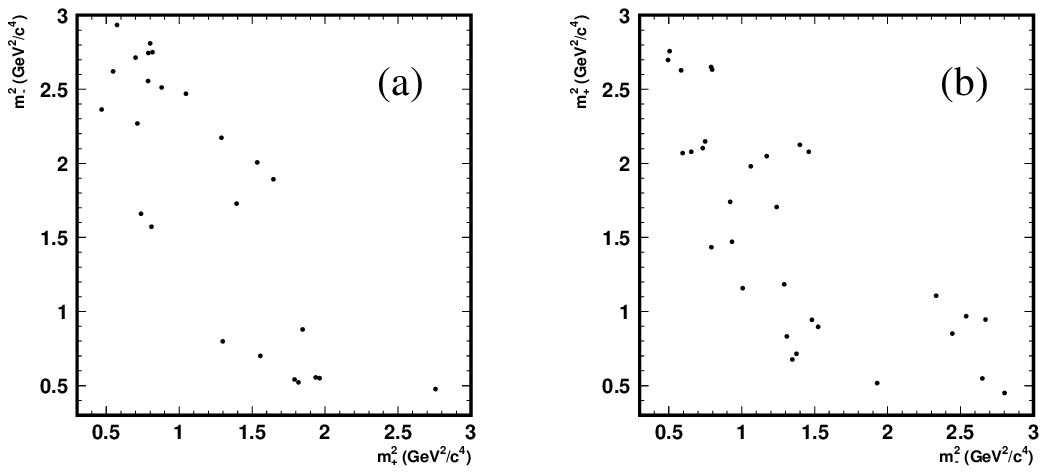,width=\textwidth}
  \vspace{-1\baselineskip}
  \caption{Dalitz plots of \dtkpp\ decay from (a) $B^+\to \tilde{D} K^{*+}$
  and (b) $B^-\to \tilde{D} K^{*-}$.}
  \label{b2dks_plots}
\end{figure}

The Dalitz plot distributions for the \dtkpp\ decay from \bdtks\ are shown in 
Fig.~\ref{b2dks_plots}.

The Dalitz plot fit procedure is similar to that used in our 
previous analysis \cite{belle_new}. The model of \dkpp\ decay 
is based on the 253 fb$^{-1}$ sample of continuum \dsdpi\ decays
and is also the same as in our \bddsk\ analysis. 

We consider three sources of background (see Table~\ref{bck_table}), and  
determine the fraction and Dalitz plot shape for each component. 
The largest contribution comes from two kinds of continuum events:
random combinations of tracks, and correctly
reconstructed neutral $D$ mesons combined with a random $K_S$ and a pion
(or a random $K^{*\pm}$).
The fraction of candidates for these events is estimated
to be $18.0\pm 4.5$\% using an event sample 
in which we make requirements that
primarily select continuum events but reject $B\bar{B}$ events. 
The shapes of their Dalitz plot distributions are 
parameterized by a third-order polynomial in the variables $m^2_+$ and 
$m^2_-$ for the combinatorial background component and a sum of $D^0$ 
and $\bar{D^0}$ shapes for real neutral $D$ mesons combined with random
$K_S$ and a pion. 

\begin{table}
\caption{Fractions of different background sources.}
\label{bck_table}
\begin{tabular}{|l|c|c|} \hline
Background source                     & Fraction \\ \hline
$q\bar{q}$ combinatorial              & $18.0\pm 4.5$\%   \\
$B\bar{B}$ events other than $B^{\pm}\to DK^{*\pm}$ 
                                      & $9.0\pm 1.5$\%    \\
Combinatorics in $D^0$ decay          & $0.6\pm 0.1$\%    \\ \hline
Total                                 & $27\pm 5$\%       \\ \hline
\end{tabular}
\end{table}

The background from $B\bar{B}$ events is subdivided into two
categories. 
The $DK^{*\pm}$ combinations coming from the decay of $D^{(*)}$ 
from one $B$ meson and $K_S$ and $\pi^{\pm}$ from the other $B$ decay 
constitute the largest part of the $B\bar{B}$ background. 
The fraction of this source is estimated to be $9.0\pm 1.5$\% using a MC study. 
The \bdks\ events where one of the neutral $D$ meson decay products is 
combined with a random kaon or pion were studied using a signal MC data set.
The estimated background fraction is $0.6\pm 0.1$\%. 
The Dalitz plot shape for both of these backgrounds is parameterized
by a sum of linear functions of variables $m^2_+$ and $m^2_-$ 
and a $D^0$ amplitude. 

\section{Results}

\begin{figure}[!htb]
  \centering
  \vspace{-1\baselineskip}
  \epsfig{figure=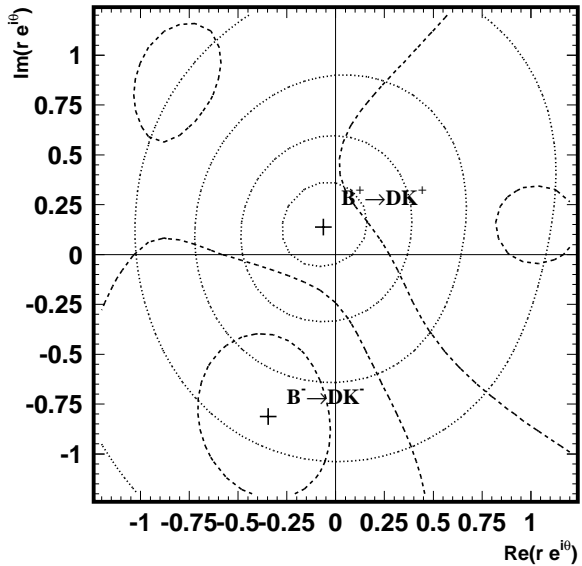,width=0.5\textwidth}
  \vspace{-1\baselineskip}
  \caption{Constraint plot of the complex amplitude ratio
           $re^{i\theta}$ for \bdtks\ decay. 
	   Contours indicate integer multiples of the standard deviation.
           Dotted contours are from $B^+$ data, dashed 
           contours are from $B^-$ data.}
  \label{compl_constr}
\end{figure}

Fig.~\ref{compl_constr} shows the constraints on the complex 
amplitude ratio $r e^{i\theta}$ separately for $B^+$ and $B^-$
samples. The global minima of the likelihood function are denoted by 
the crosses.  

A combined unbinned maximum likelihood fit to the 
$B^+$ and $B^-$ samples with 
$r$, $\phi_3$ and $\delta$ as free parameters yields the following values: 
$r=0.37\pm 0.18$, $\phi_3=112^{\circ}\pm 32^{\circ}$, 
$\delta=353^{\circ}\pm 32^{\circ}$.
The errors quoted here are obtained from the likelihood fit.
These errors are a good representation of the statistical uncertainties for
a Gaussian likelihood distribution, however in our case
the distributions are highly non-Gaussian. In addition, the errors
for the strong and weak phases depend on the values of the
amplitude ratio $r$ ({\it e.g.} for $r=0$ there is 
no sensitivity to the phases). 

\begin{figure}
  \vspace{-1\baselineskip}
  \epsfig{figure=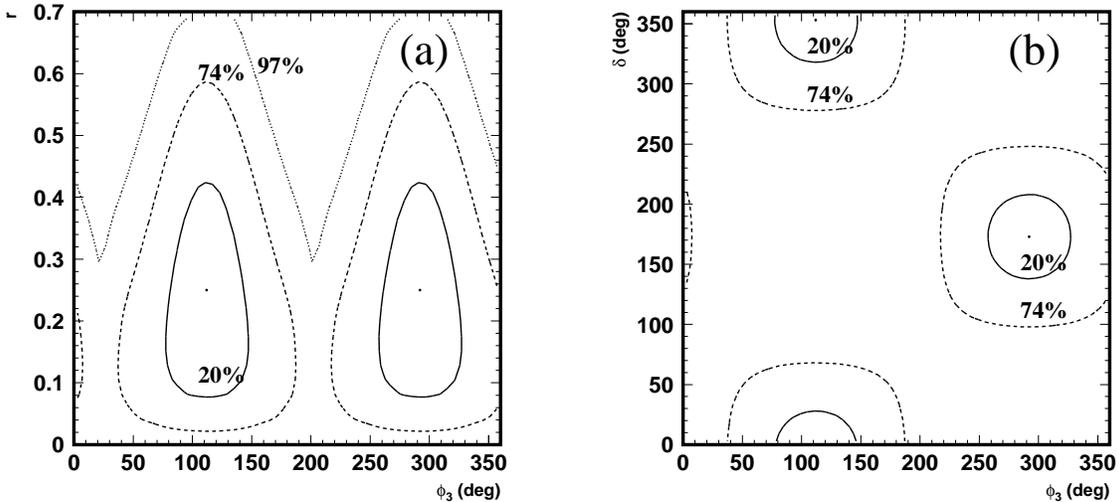,width=\textwidth}
  \vspace{-1\baselineskip}
  \caption{Confidence regions for the pairs of parameters (a) ($r$, $\phi_3$) 
           and (b) ($\phi_3, \delta$) for the \bdtks\ sample.}
  \label{b2dks_neum}
\end{figure}

As in our \bddsk\ analysis, we use a frequentist technique to evaluate the 
statistical significance of the $\phi_3$ measurement and to correct 
for the bias of the fit procedure. We use toy 
MC pseudo-experiments to obtain the probability density function (PDF)
of the fitted parameters as a function of the true parameters, 
followed by a confidence level calculation using the Neyman
method. The confidence regions for the pairs of parameters 
$(\phi_3, r)$ and $(\phi_3, \delta)$ are shown in Fig.~\ref{b2dks_neum}.
They are the projections of the corresponding 
confidence regions in the three-dimensional parameter space. 
We show the 20\%, 74\% and 97\% confidence level regions, 
which correspond to 
one, two, and three standard deviations for a three-dimensional Gaussian
distribution.
While the $\phi_3$ and $\delta$ values that are determined
from the toy MC are consistent with those that are determined in the unbinned 
maximum likelihood fits for both $\tilde{D} K^\pm$ and $\tilde{D^*}
K^\pm$, the corresponding $r$ values are significantly different. This
is caused by a bias in the unbinned maximum likelihood. Since $r$ is a
positive-definite quantity, the fit tends to return a larger value for
$r$ than its true value, particularly when $r$ is small. 

There are potential sources of systematic
error such as uncertainties in the background Dalitz plot density, 
efficiency variations over the phase space, $m^2_{\pi\pi}$ resolution, 
and possible fit biases. These are listed in Table~\ref{syst_table}. 
The effect of background Dalitz plot density 
is estimated by extracting the background shape parameters  
from the $M_D$ sidebands and by using a flat background distribution.
The maximum deviation of the fit parameters from the ``standard''
background parameterization is assigned as the corresponding
systematic error. The effect of the uncertainty in the background 
fraction is studied by varying the background
fraction by one standard deviation.
The efficiency shape and $m^2_{\pi\pi}$ resolution 
are extracted from the MC simulations. 
To estimate their contributions to the systematic error, we 
repeat the fit using a flat efficiency and a fit model that
does not take the resolution into account, respectively. 
There is no obvious control sample for the 
mode \bdks\ (with $K^{*\pm}\to K_S\pi^{\pm}$)
like \bddspi\ in the case 
of \bddsk\ modes, but we include the effect of the possible 
bias of the \bdpi\ control sample
into the systematic error to account for possible deficiency in the 
$D^0$ model, or other fit biases.

\begin{table}
\caption{Contributions to the experimental systematic error.}
\label{syst_table}
\begin{tabular}{|l|c|c|c|c|c|c|} \hline
Source               & $\Delta r$ & $\Delta\phi_3$ ($^{\circ}$) &
$\Delta\delta$($^{\circ}$)  \\ \hline
Background shape     & 0.027      & 5.7                & 4.1    \\
Background fraction  & 0.006      & 0.2                & 1.0    \\
Efficiency shape     & 0.012      & 4.9                & 2.4    \\
$m^2_{\pi\pi}$ resolution & 0.002 & 0.3                & 0.3    \\
Control sample bias  & 0.004      & 10.2               & 10.2   \\ \hline
Total                & 0.03       & 13                 & 11     \\ \hline
\end{tabular}
\end{table}

The uncertainty due to the model of $D^0$ decay is taken to be the same 
as for the \bddsk\ processes. However, the analysis of \bdks\ 
has an additional uncertainty due to the possible presence of the 
nonresonant \bdksnr\ component, which can also be considered 
as a model uncertainty. Since the nonresonant decay is described by 
the same set of diagrams as \bdks\, 
a similar $CP$ violating effect should take place 
but, in general, with different $r$ and $\delta$ from the resonant
mode. 
Thus, for the $\phi_3$ measurement
using \bdks\ mode alone, the contribution of \bdksnr\ decay can bias the 
fit parameters. To estimate the corresponding systematic uncertainty, 
we set an upper limit on the \bdksnr\ fraction, and perform a toy MC simulation 
with nonresonant contribution added to determine the fit bias. 

To limit the fraction of the nonresonant component we 
search for a \bdksnr\ signal in the sidebands of the $K^*$ invariant mass 
distribution. Since a specific final state of the neutral $D$ meson 
is not needed for this study, we combine several $D$ decay modes 
($K\pi$, $K_S\pi\pi$, $K\pi\pi^0$, $K\pi\pi\pi$) to increase the statistics. 
No requirements on event shape parameters or $K^*$ helicity are applied. 
This allows to obtain a tighter upper limit on the 
\bdksnr\ contribution due to higher signal efficiency. 

The following ranges of $K^*$ invariant mass $m_{K_S\pi}$ are 
analysed: $|m_{K_S\pi}-M_{K^*}|<50$ MeV/$c^2$ for \bdks\ events and 
$80$~MeV/$c^2<|m_{K_S\pi}-M_{K^*}|<300$ MeV/$c^2$
for \bdksnr. The number of signal events is extracted from the fit 
to $\Delta E$ distribution. The fit function is a combination of a 
Gaussian peak for the signal and a linear function for the background. 
The number of events in the $K^*$ sideband is $3\pm 40$. The number 
of events in the $K^*$ signal region is $321\pm 31$. Taking into account the 
shape of $K_S\pi$ invariant mass distribution, we obtain a \bdksnr\ 
fraction in our signal region of $0.3\pm 3.7$\%. 
The 95\% confidence level (CL) upper limit on this fraction is 6.3\%, 
which we use for estimation of the systematic error due to the 
nonresonant contribution.

To study this effect, the Dalitz plot of the $\tilde{D}$ decay is 
generated according to a decay amplitude with a 
component having different values of $r$ and $\delta$, corresponding 
to the \bdksnr\ contribution, added coherently. 
The resulting Dalitz plot distribution is then fitted with the 
standard technique, and the biases of the fit parameters $r$, $\phi_3$
and $\delta$ are obtained. The fits are performed for different values of 
$r$ and $\delta$ for the nonresonant component and for different values of the 
relative phase between \bdks\ and \bdksnr\ amplitudes. The maximum 
bias of the fit parameters is taken as a corresponding systematic error.
We obtain the following estimates of the uncertainty due to the \bdksnr\ 
contribution: $\Delta r=0.084$, $\Delta\phi_3=8.3^{\circ}$, 
$\Delta\delta=49.3^{\circ}$. We quote these errors separately from the 
other systematic uncertainties. The $\phi_3$ bias is significantly smaller
than that for the strong phase $\delta$, since $\phi_3$ is obtained 
from a difference of the total phases for $B^+$ and $B^-$ decays, and
a part of the bias cancels in this case. 

For the final results, we use the central values that are obtained by 
maximizing the PDF of the fitted parameters
and the statistical errors corresponding to the 20\% 
confidence region (one standard deviation). Of the two possible 
solutions ($\phi_3$, $\delta$ and $\phi_3+180^{\circ}$, $\delta+180^{\circ}$) 
we choose the one with $0^{\circ}<\phi_3<180^{\circ}$. The final results are 
\begin{equation*} 
r = 0.25^{+0.17}_{-0.18} \pm 0.09 \pm 0.04 \pm 0.08,~
\phi_3=112^{\circ} \pm 35^{\circ} \pm 9^{\circ} \pm 11^{\circ}\pm 8^{\circ},~ 
\delta=353^{\circ} \pm 35^{\circ} \pm 8^{\circ} \pm 21^{\circ}\pm 49^{\circ}.
\end{equation*}
The first error is statistical, the second is experimental systematics, 
the third is model uncertainty and the fourth is the error due to possible 
\bdksnr\ contribution. 
The two standard deviation interval including the 
systematic and model uncertainties is $34^{\circ}<\phi_3<190^{\circ}$. 
The statistical significance of $CP$ violation is 63\%. 

\section{Conclusion}

We report results of a measurement of the unitarity
triangle angle $\phi_3$ that uses a method based on a 
Dalitz plot analysis of the three-body $D$ decay in the process 
\bdks.
The measurement is based on a 253 fb$^{-1}$ data sample collected by 
the Belle detector. We obtain the value of 
$\phi_3=112^{\circ}\pm 35^{\circ}\pm 9^{\circ}\pm 11^{\circ} \pm 8^{\circ}$
(solution with $0^{\circ}<\phi_3<180^{\circ}$).
The first error is statistical, the second is experimental systematics, 
the third is model uncertainty and the fourth is the error due to possible 
\bdksnr\ contribution. 
The statistical significance of $CP$ violation for the combined 
measurement is 63\%. 
The method allows us to obtain a value of the 
amplitude ratio $r$, which can be used in other $\phi_3$ 
measurements. We obtain $r=0.25^{+0.17}_{-0.18}\pm 0.09\pm 0.04\pm 0.08$.

\section*{Acknowledgments}
We thank the KEKB group for the excellent operation of the
accelerator, the KEK cryogenics group for the efficient
operation of the solenoid, and the KEK computer group and
the National Institute of Informatics for valuable computing
and Super-SINET network support. We acknowledge support from
the Ministry of Education, Culture, Sports, Science, and
Technology of Japan and the Japan Society for the Promotion
of Science; the Australian Research Council and the
Australian Department of Education, Science and Training;
the National Science Foundation of China under contract
No.~10175071; the Department of Science and Technology of
India; the BK21 program of the Ministry of Education of
Korea and the CHEP SRC program of the Korea Science and
Engineering Foundation; the Polish State Committee for
Scientific Research under contract No.~2P03B 01324; the
Ministry of Science and Technology of the Russian
Federation; the Ministry of Higher Education, Science and Technology of the Republic of Slovenia;  the Swiss National Science Foundation; the National Science Council and
the Ministry of Education of Taiwan; and the U.S.\
Department of Energy.

\end{document}